\documentclass[twocolumn]{jpsj2}
\usepackage{color}

\def\runauthor{M. {\sc Miyazaki}, {\it et al.} }

\title{Variational Monte Carlo Study of a Spinless Fermion 
  $t$-$V$ Model \\ on a Triangular Lattice : 
  Formation of a Pinball Liquid
}
\author{Mitake {\sc Miyazaki}, Chisa {\sc Hotta}$^{1,2}$, 
  Shin {\sc Miyahara}$^{1,3}$, Keisuke {\sc Matsuda}$^1$ and
  Nobuo {\sc Furukawa}$^{1,3}$}
\inst{Hakodate National College of Technology, 
  Tokura-cho 14-1, Hakodate, Hokkaido 042-8501, Japan\\
  $^1$Department of Physics and Mathematics, Aoyama Gakuin University, \\
  Fuchinobe 5-10-1, Sagamihara, Kanagawa 229-8558, Japan\\
  $^2$ Kyoto Sangyo University, 
  Motoyama, Kamigamo, Kita-ku, Kyoto 603-8555, Japan\\
  $^{3}$Multiferroics Project (MF), ERATO,
  Japan Science and Technology Agency (JST), \\
  c/o Department of Applied Physics, The University of Tokyo, 
  7-3-1 Hongo, Bunkyo-ku, Tokyo 113-8656, Japan}
\recdate{\today}

\abst{
  We analyze a model of spinless fermions on a triangular lattice at half-filling interacting 
  via strong nearest-neighbor repulsive interactions, $V$, using the variational 
  Monte Carlo simulation technique. 
  The existence of three-sublattice long-range order is confirmed by 
  the finite-size scaling analysis of the charge structural factor at $V_c/t \gtrsim 12$. 
  This ordered phase shows characteristics expected for a so called ``pinball liquid" state, 
  which has the spontaneous separation of fermionic degrees of freedom into coexisting 
  Wigner crystal-like charge order (pin) and a metal (ball). 
  The pins are fixed in order to maximize the kinetic energy gain of balls which move almost freely. 
  The Fermi surface is reconstructed at $V=V_c$ and remains towards the strong coupling limit. 
  These features reminiscent of the strong correlation together with the large value of $V_c/t$ distinguishes 
  the pinball liquid from the conventional charge-density-wave. 
}

\kword{
  triangular lattice, honeycomb lattice, supersolid, 
  organic solid, charge ordering, variational Monte Carlo method, geometrical frustration 
}

\begin{document}

\sloppy
\maketitle
%XXXXXXXXXXXXXXXXXXXXXXXXXXXXXXXXXXXXXXXXXXXXXXXXXXXXXXXXXXXXXXX
%*%*%*%*%*%*%*%*%*%*%*%*%*%*%*%*%*%*%*%*%*%*
%*%*%*%*%*%*%*%*%*%*%*%*%*%*%*%*%*%*%*%*%*%*
\section{Introduction}
Geometrically frustrated lattices can provide particular situations in which strong correlation destroys conventional classical orders and sometimes gives rise to another exotic states.  Some simple examples are found in frustrated quantum spin systems, e.g. spin-gapped singlet states~\cite{okamoto92,ueda96,miyahara99}, which are realized in realistic quasi low-dimensional materials such as
 ${\rm CaV_4O_9}$\cite{taniguchi95} and ${\rm SrCu_2(BO_3)_2}$\cite{kageyama99,miyahara03}. Another recent example is the nematic order which has been observed in spin systems on a square lattice with added ring exchange or frustrated next-nearest-neighbor interactions\cite{lauhili,nic}. 
The quantum $S=1/2$ spin systems are often described by the hard core bosonic model. The model is studied intensively and the supersolidity is realized under the frustrated interactions on the triangular and the square lattices\cite{wessel05,heidarian05,melko05,mila08}. 
\par
The present paper deals with the spinless fermionic system on the triangular lattice. 
The basic nature of the ground state of this model is already studied by part of the authors\cite{hotta06, hotta06-2}; a partially charge ordered liquid called a ``pinball liquid" is proposed based on the finite-size cluster calculation~\cite{hotta06}. 
This state shows a three-sublattice structure, in which the carriers on one sublattice are localized as a Wigner-crystal (pin) and the remainders form a liquid (ball) itinerating along the other two sublattices. 
When we interpret the pin and ball as a solid and superfluid of the supersolid on the hard core bosonic model, we find that the pinball liquid has the same origin to the supersolid, i.e., order by disorder mechanism\cite{hotta06}. This suggests that although the statistics of bosons and fermions differ in two-dimension, the way they correlate is similar.

However, comprehensive analysis of the fermionic system in two-dimension is
 far difficult from the case of bosons which is already known to have a long-range order in both the solid and the liquid part\cite{wessel05}. 
Therefore the details of the fermionic ground state nature still remain unclear and several questions arise, e.g. whether there is a long-range order of solid or not, what point differs from the conventional charge-density-wave (CDW), what kind of metallic state we have (Fermi liquid or non-Fermi liquid), or even whether the metallicity remains stable in the bulk limit. 
To clarify these issues, we adopt the variational Mote Carlo (VMC) method to the half-filled ground state of the strongly interacting fermionic model on a triangular lattice. A charge correlation function indicates the realization of three-sublattice type long-range order, where two of the three sites are equivalent. At the same time, the existence of the Fermi surface is suggested in the momentum distribution function. 
\par
We shall mention that the topic is closely related to the experimental reports on organic solid, $\theta$-ET$_2X$, where a non-linear conductivity and many different kinds of short-range diffuse spots are reported in the low temperature metallic state\cite{sawano05,mori98,yamaguchi}.
This organic system is often studied in terms of the extended Hubbard model (EHM) with interactions between charges populating on the same site ($U$) and on neighboring sites ($V$)~\cite{seo04,merino05,kaneko06,watanabe06,nishimoto08}. 
In such cases further difficulty arises in analyzing this model due to large degrees of freedom including spins. 
In fact, there remains some discrepancies in the phase diagram, 
e.g. the exact diagonalization studies show that the system is a ``charge liquid"\cite{seo04} without any 
long-range order in analogue to the spin liquid. 
In contrast, the VMC and the density matrix renormalization group (DMRG) study shows that the charge degrees of freedom has translationally broken structure 
in the metallic phase at $V\sim V' \gtrsim 3t$\cite{watanabe06,nishimoto08}. 
We notice the fact that the strong coupling limit of the EHM without spin degrees of freedom can share the same classical nature as the present $t$-$V$ model at half-filling we deal with. Therefore, our results can give a clue to understand the quantum fluctuation effect of charge degrees of freedom from that Ising limit of the EHM. 
Comparison of the results of the two models will be given in $\S$4. 
%
%*%*%*%*%*%*%*%*%*%*%*
%*%*%*% fig1  %*%*%*%*
%*%*%*%*%*%*%*%*%*%*%*
\begin{figure}[t]
  \begin{center}
    \includegraphics[width=4cm]{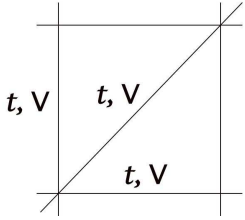}
    \caption{ 
      The lattice structure and the transfer integrals $t$ and 
      the Coulomb repulsions $V$. }
    \label{fig1}
  \end{center}    
\end{figure}
%*%*%*%*%*%*%*%*%*%*%*
%
%*%*%*%*%*%*%*%*%*%*%*%*%*%*%*%*%*%*%*%*%*%*
%*%*%*%*%*%*%*%*%*%*%*%*%*%*%*%*%*%*%*%*%*%*
\section{Formulation}
%\noindent
The $t$-$V$ model of spinless fermions is given explicitly as, 
\begin{eqnarray}
  {\cal H} & = & {\cal H}_t+{\cal H}_V  \nonumber \\
  & = & \sum_{\langle i,j\rangle}(-t c_{i}^{\dagger}c_{j}+{\rm h.c.}
  +V n_{i}n_{j}).  \label{eq:hamil}
\end{eqnarray}
Here, $c_{i}^{\dagger}$ ($c_{i}$) is the creation (annihilation) operator 
of fermions at site $i$ and $n_{i}=c^{\dagger}_{i}c_{i}$. 
We allow a hopping between nearest-neighbors as well as
next-nearest-neighboring sites along only one diagonal direction
on a square lattice with a hopping amplitude $t$ ($t > 0$)
as shown in Fig.~\ref{fig1}.
The hopping integral $t$ is taken as the unit of energy and 
$V$ is the near-neighbor Coulomb repulsion energy. 
The sum on $\langle i,j\rangle$ runs over pairs 
of neighboring sites. 
Then, the dispersion is given by 
\begin{equation}
  E({\mib k})=-2t\cos(k_x)-2t\cos(k_y)-2t\cos(k_x+k_y).
\end{equation}
This model is equivalent to the equilateral triangular lattice. 
Here, we remark for the half-filled case in order to clarify the features 
of the pinball liquid state in the strongest geometrical frustration. 
The similar analysis for other filling, focusing on 
Na$_2$CoO$_4$, was reported in ref.~\citen{Motrunich04}. 
\par
To investigate ground state with strong interactions, 
we compute the variational energy,
$E={\langle \Psi |H|\Psi\rangle}/{\langle\Psi|\Psi\rangle}$,
by using the VMC method. Since the Hamiltonian consists only of the 
near-neighbor coupling term, it is necessary to add a 
Jastrow~\cite{jastrow55} term to the wave function to lower the  
variational energy: we study the intersite correlation effect  by using 
Jastrow-type trial wave function $|\Psi_T\rangle$ as defined by
\begin{equation}
  |\Psi_T\rangle=P_h|\phi\rangle,
\end{equation}
where $|\phi\rangle$ is an one-body wave function determined by 
diagonalizing the mean-field Hamiltonian.  $P_h$ is a 
near-neighbor projection operator given by 
\begin{equation}
P_h=\prod_{\langle ij\rangle}h^{n_{i}n_{j}},
\end{equation}
where $h$ is a variational parameter which controls the weight of  
the wave function; when the fermion exists on
both $i$ and $j$ sites, the wave function is multiplied by $h$,
while it remains unchanged for the other cases.
In the limit of $h=1$, the wave function corresponds to $|\phi\rangle$ itself, while at $h\rightarrow0$ the occupation on nearest-neighbor site 
is completely suppressed. 

In the present paper, 
we adopt the $L\times L$ ($L=6 \sim 18$) lattices with periodic 
boundary conditions in both directions, and optimize the energy expectation values 
for different $V$ with the total Monte Carlo step greater than $3\times10^7$. 
The system does not satisfy the closed shell condition for half-filling. 
However, we confirmed that there is no qualitative difference in the 
result of closed shell with periodic-antiperiodic boundary condition. 
We choose two different kinds of one-body wave functions, 
$|\phi\rangle=|\phi_{\rm FS}\rangle$ and $|\phi_{\rm CDW}\rangle$, which are 
the Fermi sea and the CDW solution of the mean-field calculation, 
respectively. 
The former has a translational symmetry and gives 
the simplest and unbiased results when projected. 
We also examine the possibility of the breaking of the translational 
symmetry by adopting the latter CDW wave function. 
The comparison of these two cases gives us useful information. 
%*%*%*%*%*%*%*%*%*%*%*%*%*%*%*%*%*%*%*%*%*%*
%*%*%*%*%*%*%*%*%*%*%*%*%*%*%*%*%*%*%*%*%*%*
\section{Results}
\subsection{Projected Fermi sea}
\begin{figure}[t]
  \begin{center}
    \includegraphics[width=7.5cm]{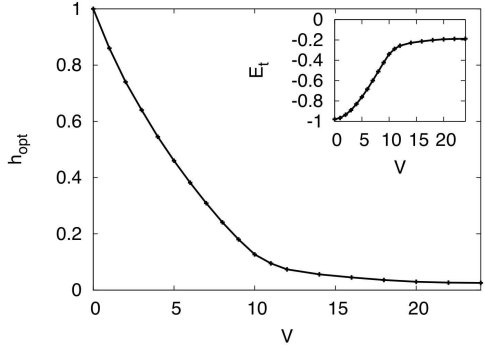}
    \caption{The optimized near-neighbor correlation factor 
      $h_{\rm opt}$ as a function $V$ by using the Jastrow type 
      wave function based on the Fermi sea. The system is $12\times 12$.
      The inset shows the expectation value $E_{\rm opt}$ 
      of ${\cal H}_t$ part as a function of $V$.}
    \label{fig2}
  \end{center}
\end{figure}

%\noindent
First, we assume the projected Fermi sea, $P_h|\phi_{\rm FS}\rangle$, 
as a variational wave function for the normal phase. 
The $V$-dependence of the optimized variational parameter $h_{\rm opt}$ is shown 
in Fig.~\ref{fig2}. The $h_{\rm opt}$ rapidly decreases from the non-interacting value, 
$h_{\rm opt}=1$, as $V$ increases. 
The functional form changes at $V \sim 12$, which suggests a change of the
character of the state.
The expectation values of kinetic part ${\cal H}_t$ 
of eq. (\ref{eq:hamil}) is also shown in the inset as a function $V$. 
It shows a nearly $V$-independent insulating behavior at $V > 12$, however, 
even in this region there is a finite kinetic energy gain. 
%
%*%*%*%*%*%*%*%*%*%*%*
%*%*%*% fig2  %*%*%*%*
%*%*%*%*%*%*%*%*%*%*%*
%*%*%*%*%*%*%*%*%*%*%*
\par
\begin{figure}[t]
  \begin{center}
    \includegraphics[width=7.5cm]{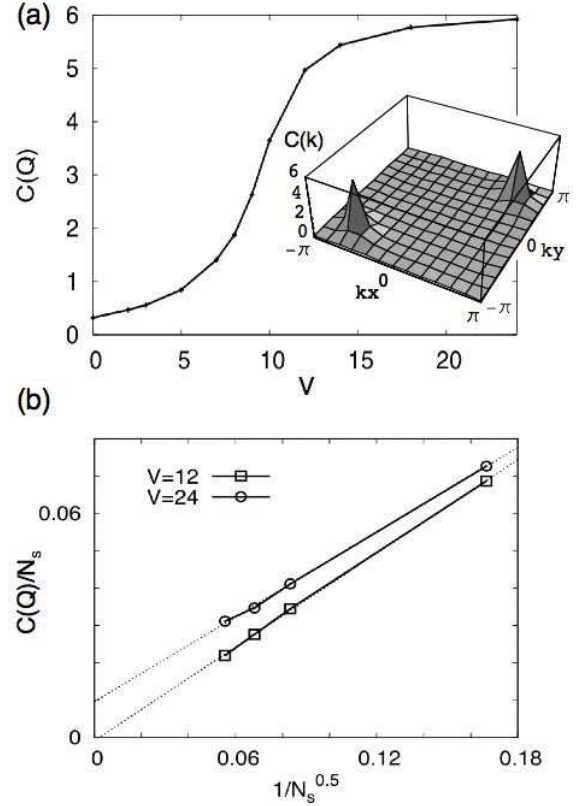}
    \caption{(a) Intensity of the peak of the charge correlation function $C({\mib k})$
      at ${\mib k}={\mib Q}=\pm (2\pi/3, 2\pi/3)$ as a function of $V$. 
      The trial wave function, $P_h$$|\phi_{\rm FS}\rangle$, is adopted on the $N_s=12 \times 12$ lattice. 
      The inset shows the $k$-dependence of  $C({\mib k})$ at $V=12$. 
      (b) System-size dependence of charge structure 
      factors $C({\mib Q})$ for $V=12$ and $V=24$.
      The dotted lines are least-squares fits to the data.}
    \label{fig3}
  \end{center}
\end{figure}

Next, to investigate the instability towards charge order, 
we measure the charge correlation function by the optimal wave function, which is defined as
\begin{equation}
  C({\mib k}) = \frac{1}{N_s}\sum_{i,j}
  \left(\langle n_{i}n_{j}\rangle-\langle n_{i}\rangle\langle n_j\rangle\right)
       {\rm e}^{{\rm i}({\mib r_i}-{\mib r_j})\cdot {\mib k}},
\end{equation}
where $N_s$($=L\times L$) is the number of sites. 
In inset of Fig.~\ref{fig3}(a), two peaks are observed in $C({\mib k})$ 
at ${\mib k}= \pm (2\pi/3, 2\pi/3)$ for $V=12$. 
This wave number corresponds to the charge modulation patterns based on a three-sublattice structure. 
As shown in Fig.~\ref{fig3}(a), the peak in $C({\mib k})$ grows as $V$ increases 
and starts to saturates above the inflexion point, $V \gtrsim 10$. 
\par
To see whether the state has a long-range order, 
we perform a finite-size scaling analysis and examine whether the charge structure factor 
$C({\mib Q})$ diverges in the thermodynamic limit. 
If the long-range order exists, the charge structure factor is expected to behave as 
\begin{eqnarray}
C({\mib Q})=N_s\Delta^2+O \left( \sqrt{N_s} \right), 
\end{eqnarray}
where we expect $C({\mib Q})/N_s$ to follow a linear function of $1/\sqrt{N_s}$ 
with a nonzero intercept which gives the bulk order parameter $\Delta$. 
We plot $C({\mib Q})/N_s$ versus $1/\sqrt{N_s}$ in Fig. 3(b). 
The nonzero extrapolation value appears when $V \gtrsim 12$, which implies the 
existence of a long-range order based on the three-sublattice structure above 
the phase transition point, $V_c \simeq 12$\cite{miyazaki}. 
\par
We shall emphasize that in this scaling analysis the wave function has translational symmetry, 
so that $V_c$ gives a reliable value in the bulk limit. 
Actually, $V_c$  has the same order of magnitudes as the onset value of the supersolid order in the hard core bosonic $t$-$V$ model by the quantum Monte Carlo study 
\cite{wessel05}, $t/V_c \sim 0.12$, and is consistent. 
If we consider instead the projected CDW wave function with three-fold periodicity 
(i.e., translational symmetry is broken by hand), 
the value of $V_{\rm CDW}$ shall be several times smaller as we show in the next section. 
\par
%*%*%*%*%*%*%*%*%*%*%*
%*%*%*% fig3  %*%*%*%*
%*%*%*%*%*%*%*%*%*%*%*
%*%*%*%*%*%*%*%*%*%*%*
\par
%% 
%*%*%*%*%*%*%*%*%*%*%*%*%*%*%*%*%*%*%*%*%*%*
%*%*%*%*%*%*%*%*%*%*%*%*%*%*%*%*%*%*%*%*%*%*
\subsection{Projected charge-density-wave}
%\noindent
In the presence of the long-range order, the translational symmetry is broken in the bulk limit. 
However, in the present finite-size cluster, the projected Fermi sea wave function has 
the translational symmetry, and the orders appear 
as the correlation which lasts towards the bulk limit. 
There shall instead be a symmetry-broken wave function that could give better description 
to the bulk charge ordering. 
As such a candidate, we employ another projected wave function, $P_h|\phi_{\rm CDW}\rangle$, 
where $|\phi_{\rm CDW}\rangle$ is obtained as a CDW solution of a mean-field Hamiltonian, 
\begin{eqnarray}
 && {\cal H}_{\rm MF}=\sum_{\langle i,j\rangle}(-t c_{i}^{\dagger}c_{j}+{\rm h.c.})\nonumber\\
  &&+\sum_i \big( \Delta_{1} \cos\left({\mib Q}\cdot{\mib r}_i\right)
  + \Delta_{2} \sin\left({\mib Q}\cdot{\mib r}_i\right) \big) n_i,\ \ \ \
\end{eqnarray}
where $\Delta_1$ and $\Delta_2$ are CDW order parameters and ${\mib Q}= \pm(2\pi/3, 2\pi/3)$ is a wave vector which characterizes the state. 

\begin{figure}[t]
  \begin{center}
   \includegraphics[width=8.5cm]{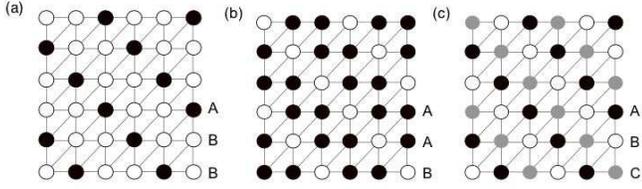}
    \caption{Representative charge ordered patterns studied in this paper. 
      Charge density of (a) A-B-B type, (b) A-A-B type, and (c) A-B-C type. 
      Black and white circles represent charge-rich A and charge-poor B sites, respectively. 
      Gray circles denote intermediate C sites.}
    \label{fig4}
  \end{center}
\end{figure}

Charge configurations of the mean-field solutions
exhibit three different classes,
A-B-B, A-A-B and A-B-C type, as shown schematically in Figs.~\ref{fig4}, 
where A, B, and C denote the charge-rich sublattice, the charge-poor sublattice, 
and the sublattice with the intermediate charge density 
(nearly charge-neutral), respectively. 
The parameters ($\Delta_1>0$, $\Delta_2=0$) and ($\Delta_1<0$ $\Delta_2=0$) 
correspond to A-A-B and A-B-B type of structure, respectively. 
The parameters ($\Delta_1=0$ and $\Delta_2\neq 0$) realize an A-B-C type structure. 
In the VMC calculation, we take $h$ and $\Delta_1$(or $\Delta_2$) as variational parameters, 
and find the optimized solution with finite $\Delta_1>0$ at $V \ge V_{\rm CDW} \sim 3$. 
This A-A-B type solution has lower energy than the A-B-C type one. 
The energy difference is estimated as $(E_{\rm ABC}-E_{\rm AAB})/N_s=0.014$ at $V=10$ 
on a $N=12\times 12$ cluster. 
This result is different from the exact diagonalization study which shows A-B-C type of charge correlations~\cite{hotta06}, which is possibly attributed to the smaller cluster size in that previous study. 
In the following, only the charge configuration of the A-A-B type is considered. 
\par
%*%*%*%*%*%*%*%*%*%*%*
%*%*%*% fig4  %*%*%*%*
%*%*%*%*%*%*%*%*%*%*%*
%*%*%*%*%*%*%*%*%*%*%*
\par
\begin{figure}[t]
  \begin{center}
    \includegraphics[width=7.5cm]{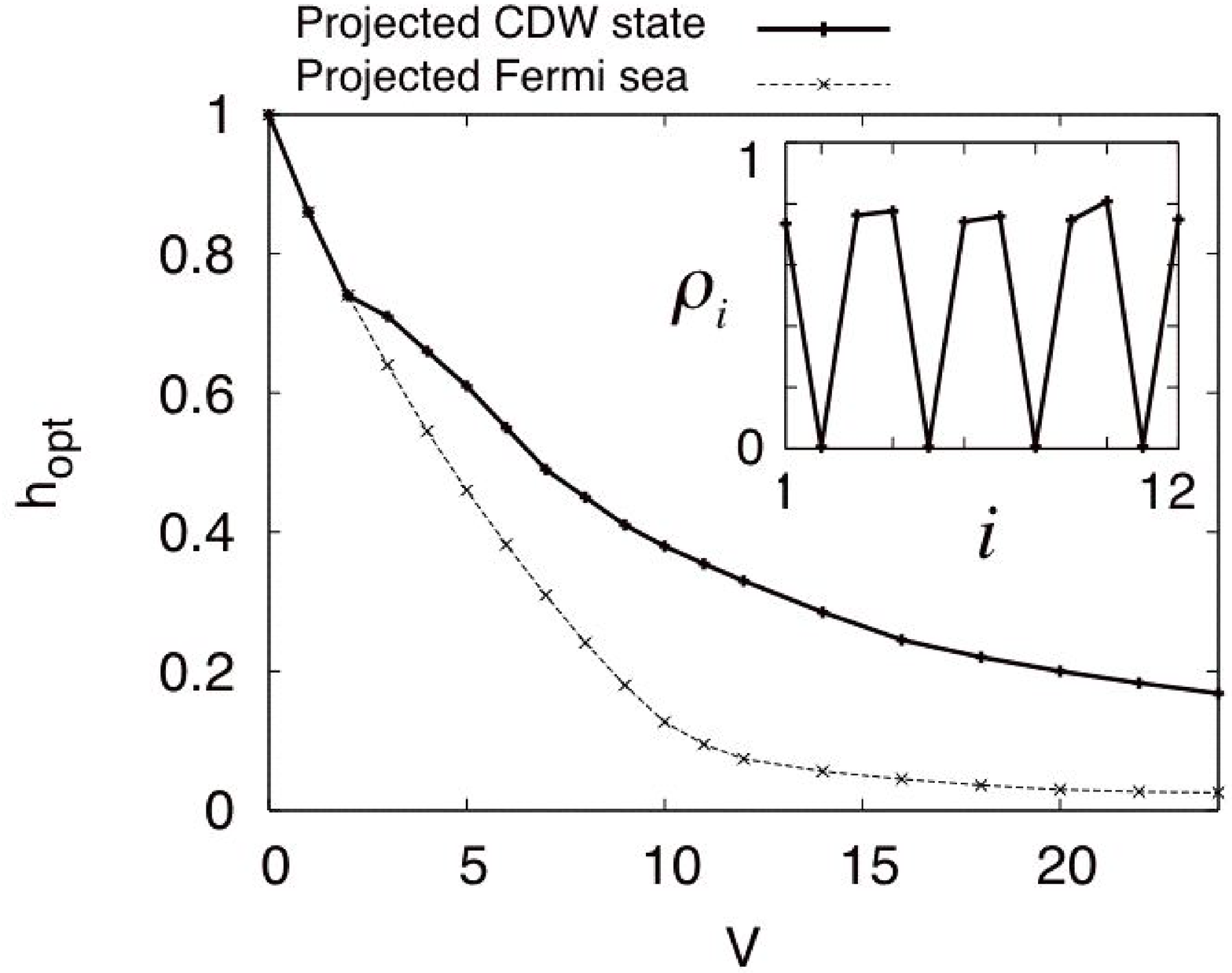}
     \caption{Optimized $h_{\rm opt}$ as a function of $V$ 
      based on the CDW trial wave function, $P_h|\phi^{\rm AAB}_{\rm CDW}\rangle$. 
      The result of fhe Fermi sea function, $P_h|\phi_{\rm FS} \rangle$, (Fig. 2) 
      is plotted here together for comparison. 
      Inset shows the expectation value of charge density along the $x$-axis on the $N_s=12\times 12$ lattices 
      at $V=24$ by the $P_h|\phi^{\rm AAB}_{\rm CDW}\rangle$. }
     \label{fig5}
  \end{center}
\end{figure}

Figure~\ref{fig5} shows the $V$-dependence of $h_{\rm opt}$ of the projected A-A-B type wave function. 
Following the emergence of finite $\Delta_1>0$ at $V > V_{\rm CDW}\sim 3$, 
$h_{\rm opt}$ shows significant deviation and the energy is lowered 
compared to those of the projected Fermi sea. 
The expectation value of the charge density, $\rho_i=\langle c_i^{\dag}c_i\rangle$, 
is given in the inset of Fig.~\ref{fig5}. 
The amplitude of pins on B-site is almost 0, which means that the holes are strongly localized as a pin. 
Such a hole-pinned configuration is expected to have the same energy with the electron-pinned configuration 
(i.e. A-B-B type with $\rho_A=1$) at half-filling if one assumes a particle-hole (p-h) symmetry. 
In the present fermionic system we always have the hole-pinned state, 
since the triangular lattice has p-h {\it asymmetry}. 
If we take the negative hopping amplitude $t<0$, an electron-pinned pinball liquid is expected instead. 
\par
The rigorous pinball liquid state is created just off the filling factors, $\rho=1/3$ and $2/3$, 
by adding a single electron or a hole, respectively, at $t/V\rightarrow 0$~\cite{hotta06}. 
Combined with the present results, we may conclude that 
the pinball liquid state is stable over the entire range of the charge density $1/3 < \rho < 2/3$. 
This phase is further classified into a hole-pinned ($\rho \sim 2/3$) and an electron-pinned ($\rho \sim 1/3$) 
ones, and the transition between them shall exists at $\rho > 1/2$. 
\par
%*%*%*%*%*%*%*%*%*%*%*
%*%*%*% fig5  %*%*%*%*
%*%*%*%*%*%*%*%*%*%*%*

%*%*%*%*%*%*%*%*%*%*%*
%*%*%*% fig6  %*%*%*%*
%*%*%*%*%*%*%*%*%*%*%*
\begin{figure}[tbp]
  \begin{center}
    \includegraphics[width=8.8cm]{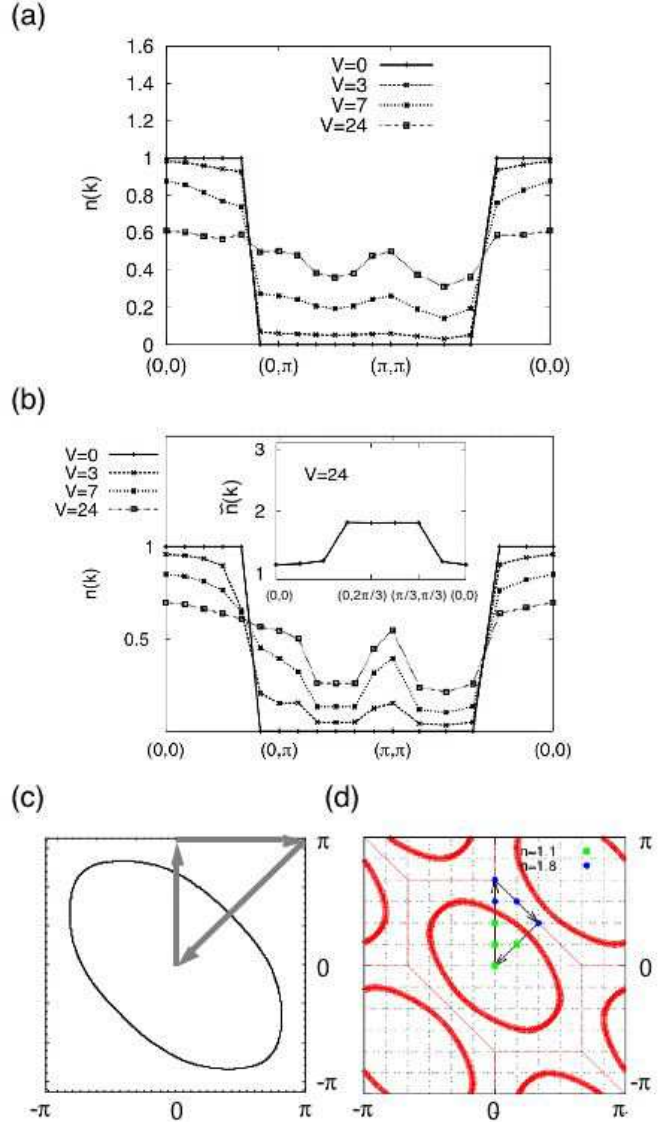}
%\vspace{-6mm}
    \caption{
      Momentum distribution functions for various $V$ on the $N_s=12\times 12$ system; 
      Panel (a) and (b) show the ones based on 
      $P_h$$|\phi_{\rm FS}\rangle$ and $P_h$$|\phi^{\rm AAB}_{\rm CDW}\rangle$, respectively, 
      which are plotted within the first Brillouin zone of the normal state, i.e. along the arrows 
      in panel (c). 
      The inset of panel (b) shows $\tilde{n}(k)$ at $V=24$, which is the replotted $n(k)$ within the folded Brillouin zone for 
      the three-sublattice, i.e. along the arrows in panel (d). 
      Panels (c) and (d) show the Fermi surfaces of an non-interacting 
      fermions on the triangular lattice at half-filling and on the honeycomb lattice at 1/4-filling, 
      respectively. In (d) the Brillouin zone folded by 1/3 from the one in (c), 
      which is shown in broken line. 
      The dots on panel (d) denote the representative data points of the momentum distribution 
      function in the inset of panel (b).
      }
    \label{fig6}
  \end{center}
\end{figure}

%*%*%*%*%*%*%*%*%*%*%*%*%*%*%*%*%*%*%*%*%*%*
%*%*%*%*%*%*%*%*%*%*%*%*%*%*%*%*%*%*%*%*%*%*
\subsection{Fermi surface reconstruction}
%\noindent
In order to understand the origin of the breaking of translational symmetry in more detail, 
we analyze the momentum distribution function at the Fermi surface,  
\begin{equation}
  n({\mib k})=\frac{1}{N_s}\sum_{i,j}\langle c^{\dagger}_{i} c_{j}\rangle 
  {\rm e}^{{\rm i}({\mib r_i}-{\mib r_j})\cdot {\mib k}}.
\end{equation}
Figures~\ref{fig6}(a) and \ref{fig6}(b) show the $V$-dependence of $n({\mib k})$ measured with 
the optimized parameters along the path $(0,0)$-$(0,\pi)$-$(\pi,\pi)$-$(0,0)$ 
in the Brillouin zone for the trial wave functions, $P_h|\phi_{\rm FS} \rangle$ and $P_h|\phi^{\rm AAB}_{\rm CDW}\rangle$, respectively. 
For comparison, we show the Fermi surface for the non-interacting case
in Fig.~\ref{fig6}(c).

In Fig.~\ref{fig6}(a) we find a large jump-like structures in $n(k)$ for  $P_h|\phi_{\rm FS} \rangle$ at $V=0,3,7$. 
The positions of the jumps are consistent with the Fermi surface
of the non-interacting system.
When $V=24$, the jump is blured and instead a structure at around $(\pi,\pi)$ appears. 
Thus, it is likely that the long-range order at $V=24$ is related to the reconstruction of the Fermi surface. 

Actually, the reconstruction distinctively takes place 
when we adopt the wave function 
$P_h|\phi^{\rm AAB}_{\rm CDW}\rangle$ at $V=24$; 
new discontinuities are found both along $(0,\pi)$-$(\pi,\pi)$ and along 
$(\pi,\pi)$-$(0,0)$ in Fig.~\ref{fig6}(b). 
It is detected more clearly by assuming the three-fold periodicity
and considering a momentum distribution function defined on the
reduced Brillouin zone,
\begin{equation}
 \tilde{n}({\mib k})=\sum_{\mib G} n(\mib k +\mib G),
\end{equation}
with $\mib G$ being the reciprocal lattice vector, 
as shown in the inset of Fig.~\ref{fig6}(b) along the path $(0,0)$-$(0,2\pi/3)$ and $(\pi/3,\pi/3)$-$(0,0)$. 
In Fig.~\ref{fig6}(d), 
we also plot the representative values of $\tilde{n}({\mib k})$ 
within the folded Brillouin zone. 

Let us focus only on A-sites in Fig.~\ref{fig4}(b).
Namely, when the hole-pinned B-sites are depleted from the triangular
lattice, remaining carriers propagate on  a honeycomb lattice formed by
A-sites.
In Fig.~\ref{fig6}(d), we also plot 
 the Fermi surface of the non-interacting system on the
honeycomb lattice at 1/4-filling. 
Note that  the honeycomb bands
have a double structure within the first Brillouin zone,
and at 1/4-filling (which corresponds to $N_s/6$-balls on $2N_s/3$-sites) 
there exist a half-filled band and an empty band so that
the large ellipsoidal Fermi surface covers half the Brillouin zone. 
The present result shows that the 
location of the discontinuity of $\tilde {n}({\mib k})$ possibly falls 
on the honeycomb Fermi surface, 
which suggests that this Fermi surface describes well the metallic state of the pinball liquid at $V > V_c$.

Therefore, it seems that the large Fermi surface at $V=0$ (Fig.~\ref{fig6}(c)) 
is discontinuously transformed to another large Fermi surface at $V>V_c$. 
At the transition, the number of carriers 
decreases by $N_s/3$ which corresponds to the number of fixed pins.
\par

The ellipsoidal Fermi surface in the non-interacting case suggests that 
the charge susceptibility $\chi_0$ only takes a broad peak structure\cite{udagawa07} 
disadvantageous of the nesting instability. 
Therefore, we expect to have large fluctuations of charges at $V\rightarrow V_c$ over a 
certain range of wave numbers around the broad peak. 
Even in such case, the mean-field solutions may allow for a phase transition into the CDW state 
which is usually considered as an artifact of underestimating the charge fluctuation. 
Since $P_h|\phi_{\rm CDW}\rangle$ assumes the periodicity by hand in a mean-field manner, 
it also overestimates the three-fold structure although the true long-range order is still absent. 

In fact, our results show that the sign of reconstruction 
is already found at $V=3$, while we still find a distinct jump near $(0,\pi)$ 
in $n({\mib k})$ in both $V=3,7$ similar to the non-interacting case,
as depicted in the main panel of Fig.~\ref{fig6}(b).
Therefore it is reasonable to consider that the Fermi surface at $V<V_c$ is close to 
the non-interacting one in Fig.~\ref{fig6}(c). 
The precursor of the pinball liquid structure appears below the true long-range order transition 
due to the short range correlation, while the true reconstruction of the Fermi surface may 
take place only at $V\ge V_c$. 
The fact that $V_c$ is four times as large as $V_{\rm CDW}$ also suggests that these two 
cannot merge and that the Fermi surface reconstruction at $V_c$ is not attributed to the nesting 
instability. 
\par
At $V=24$, the correlation effect is as strong as to have $h_{\rm opt}\sim 0.2 \ll 1$, 
while $\tilde{n}({\mib k})$ (inset of Fig.~\ref{fig6}(b)) behaves almost like a step function of the free fermions. 
This may indicate that once the pins are fixed {\it the balls move almost freely} 
in a one-body-like manner in this phase which extends towards the strong coupling limit. 
\par 

%*%*%*%*%*%*%*%*%*%*%*%*%*%*%*%*%*%*%*%*%*%*
%*%*%*%*%*%*%*%*%*%*%*%*%*%*%*%*%*%*%*%*%*%*
\subsection{Energy gain}
\begin{figure}[tbp]
  \begin{center}
    \includegraphics[width=7.6cm]{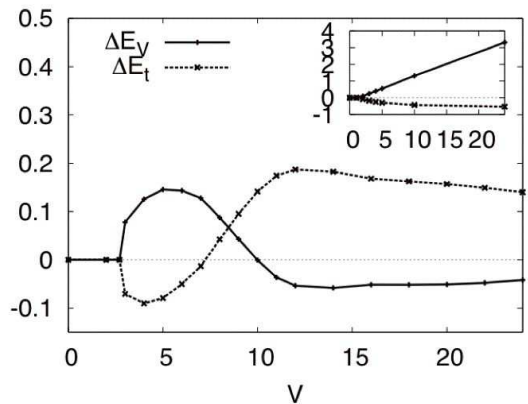}
    \caption{The kinetic and interaction parts of the optimized charge 
      condensation energy, 
      $\Delta E_t \equiv E_t(P_h|\phi_{\rm FS}\rangle)
      -E_t(P_h|\phi^{\rm AAB}_{\rm CDW}\rangle)$ and 
      $\Delta E_V \equiv E_V(P_h|\phi_{\rm FS}\rangle)
      -E_V(P_h|\phi^{\rm AAB}_{\rm CDW}\rangle)$, 
      respectively as a function of $V$. The data is from $12\times 12$ lattices.
      The inset shows the result for mean-field calculation.}
    \label{fig7}
  \end{center}
\end{figure}

%\noindent
In order to characterize the phase above the transition, 
we finally analyze the charge condensation energy, $\Delta E=E(P_h|\phi_{\rm FS}\rangle)-E(P_h|\phi^{\rm AAB}_{\rm CDW}\rangle)$, 
which is the energy gain in breaking the translational symmetry. 
The kinetic and the interaction part of the condensation energy are separately plotted as a function of 
$V$ in Fig.~\ref{fig7}. 
For comparison, we also show the condensation energy of the mean-field wave function in the inset, 
$\Delta E=E(P_h|\phi_{\rm FS}\rangle)-E(|\phi_{\rm MF}\rangle)$. 
This purely mean-field state is obtained by setting $h=1$, i.e.,$|\phi_{\rm MF}\rangle=|\phi_{\rm CDW}\rangle (h=1)$. 
The CDW transition takes place at different points, $V_{\rm CDW}\sim 3$ and $2$ 
%%$V_{\rm CDW}= 2.7$ and $1.8$ 
for the projected and purely mean-field wave functions, respectively, 
reflecting the degree of correlation included. 
Therefore, we shall again stress that the CDW phase is artifactual
and here we discuss it just for references. 
\par
In the CDW phase, $3 \lesssim V \lesssim 9$, in order to break the translational symmetry 
one shall cost the kinetic energy while gaining the interaction energy, 
i.e., $\Delta E_V>0$ and $\Delta E_t<0$. 
This mechanism is essentially the same as the case of mean-field wave function (see the inset 
of Fig.\ref{fig7}), which is a weak coupling picture. 
This suggests that the obtained projected CDW state is a mean-field like state, 
and by including the correlation effect in a more proper manner in the presence of translational symmetry, 
the kinetic energy gain shall be enhanced. 

On the other hand, at $V\gtrsim 10$, the kinetic energy gain is maximized while 
there is a cost of the interaction energy, i.e., $\Delta E_V<0$ and $\Delta E_t>0$, 
when adopting the projected CDW wave function. 
It means that although both wave functions have long-range order, the explicit breaking of translational 
symmetry (projected CDW wave function) is energetically favored due to large $E_t$; 
in the projected CDW state, fixing the pins stabilizes the metalic path (balls are less scattered), 
i.e., kinetic energy gain is maximized. 
In the projected Fermi sea the charges can fluctuate at most to avoid the interaction 
(the strong correlation effect). Such fluctuation is suppressed in the projected CDW state 
to some extent by the periodicity, which leads to the interaction energy loss. 
Thus, maximizing the kinetic energy gain plays a key role in stabilizing the long-range ordered 
pinball liquid at $V>V_c$. 

Similar characteristic behavior of $\Delta E_V$ and $\Delta E_t$ at around the phase transition point 
is found by Yokoyama {\it et al.} in the superconducting phase transition and 
order-disorder transition in the Hubbard model~\cite{yokoyama06}, 
which might be a universal features found in the strongly correlated systems. 
%*%*%*%*%*%*%*%*%*%*%*
%*%*%*% fig7  %*%*%*%*
%*%*%*%*%*%*%*%*%*%*%*

%*%*%*%*%*%*%*%*%*%*%*%*%*%*%*%*%*%*%*%*%*%*
%*%*%*%*%*%*%*%*%*%*%*%*%*%*%*%*%*%*%*%*%*%*
\section{Summary and discussion}
%\noindent
In the present paper we studied the ground state of the two-dimensional spinless fermion $t$-$V$ model 
on a triangular lattice at half-filling by the VMC method. 
We found the strong indication for the long-range order (spontaneous breaking of the translational symmetry) 
of the pinball liquid state at $V>V_c \sim 12$; 
partially localized holes (not the electrons) on one of the three-sublattices form a long-range order, while 
the other two sublattices equivalently contribute to the metallic conduction, 
i.e., an A-A-B structure which breaks the particle-hole symmetry. 
It is stabilized by suppressing the fluctuation of pins, allowing for a stable (almost free) 
metallic path to gain kinetic energy, which is consistent with the strong coupling picture\cite{hotta06}. 
By the finite-size scaling analysis based on the wave functions {\it with the translational symmetry}, 
the results give the first evidence of the existence 
of the long-range order and of the metallicity stabilized towards the bulk limit at $V\ge V_c$. 
\par
Regarding the phase at $V_{\rm CDW}<V<V_c$, we do not have a conclusive answer. 
The three-fold periodic long-range order is absent in the bulk limit. 
Therefore, the CDW wave function obtained variationally by projecting 
the mean-field solution does not give the best description of 
this phase since the translational symmetry is broken by hand. 
The search for another variational wave function without symmetry breaking 
is beyond the present scheme. Meanwhile our results suggest a short-range order 
of fluctuating three-sublattice type of structure. 
Therefore, we may safely conclude that the short-range order of charges grows with increasing $V$ due to 
strong correlation effect and at $V=V_c$, a long-range order appears. 
The reconstruction from a large Fermi surface to another large Fermi surface 
shall takes place at $V=V_c$, where the number of carriers decreases 
by the number of crystallized pins. 
\par
Let us briefly discuss the comparison with the related theoretical studies of 
the electronic system on the anisotropic triangular lattice. 
The metallic charge order is found in the EHM at quarter-filling, 
which is to be compared to the present $t$-$V$ model at half-filling; 
the exact diagonalization study mentions the idea of ``charge liquid" without long-range order 
which completely overlooked the geometrically favored three-sublattice configuration 
due to small cluster size\cite{merino05}. 
We stress that the frustration induced metallic state they discussed has {\it long-range order} 
and is to be discussed in analogy with the ``supersolid"\cite{wessel05,heidarian05,melko05} 
and not the ``spin liquid". 
Actually, it is figured out in the VMC study on EHM based on the mean-field wave function 
as the A-B-B type of order at $V_c \gtrsim 4$, while the DMRG gives A-A-B type of order at $V_c \gtrsim 3$. 
The reason for the different types of configuration in the same EHM model is unclear. 
However, the DMRG gives the reasonable hole density, $n_A\sim 1.5$ and $n_B \sim 0$, 
which is well regarded as the pinball liquid but not as the CDW (see the inset of Fig. \ref{fig5}). 
\par
As for the value of their $V_c$ (without scaling analysis) four times smaller than our estimation 
we speculate two possible reasons; 
one is that in the EHM the pins consists of the doubly occupied holes\cite{nishimoto08}, 
and at $U \ll \infty$, it may help to stabilize the long-range order. 
The magnetism which is incompatible with the pinball liquid is suppressed by this 
double occupancy. 
Another possibility is that our $V_{\rm CDW}\sim 3$ based on the mean-field CDW wave function 
corresponds to their $V_c$, which is actually close to the mean-field value. 
Therefore, in the former VMC study, the same treatment with the present study might 
give quantitatively the similar result. 
The latter DMRG might also overestimate the orders due to Friedel oscillations 
under the open boundary condition, and the true long-range order may shift to larger values of $V$. 
%*%*%*%*%*
\par
Regarding the short-range orders in the weak coupling phase, 
the DMRG study is consistent with our interpretaion\cite{nishimoto08}. 
The competition of three-sublattice and stripe type of charge ordering is discussed in more detail 
in the RPA study by including the electron-phonon(e-h) coupling terms to the EHM\cite{udagawa07}. 
The phase transition they discuss corresponds to our $V_{\rm CDW}$ and we also consider that 
it may remain as short range order if one excludes the effect of e-h coupling. 
From all these reports, we shall conclude that in both $t$-$V$ model and EHM, 
a short-range order exists in the weak coupling region which transforms 
into the long-range ordered pinball liquid at strong coupling 
although there remains some quantitative discrepancies. 
The robustness of the metallicity is guaranteed even in the presence of spin degrees of freedom 
and for systems with different types of non-interacting density of states. 
\par
The detailed comparison of the above mentioned theoretical findings and the experiments on organic $\theta$-ET$_2X$ systems\cite{sawano05,mori98,yamaguchi} are given elsewhere\cite{watanabe06,udagawa07}, and 
we just provide in this paper the essential picture of the charge degrees of freedom. 
Finally, we shall mention that not only in organic solids but in the transition metal oxide, AgNiO$_2$ \cite{coldea07} 
there exists a charge ordered metallic state which is possibly interpreted as a pinball liquid state. 
Although there is an orbital degrees of freedom in this material, the essential picture should be the same. 
We expect that more materials will be found that show such particular features of 
the strongly correlated metallic system on the similar geometrically frustrated lattices. 

\section*{Acknowledgment}
M. M acknowledges the supports from
Japan Science and Technology Agency.
This work is supported by Grant-in-Aid from the
Ministry of Education, Culture, Sports, Science and
Technology of Japan.


\section*{References}
\begin{thebibliography}{99}
\bibitem{okamoto92}
  K. Okamoto and K. Nomura:
  Phys. Lett. A \textbf{169} (1992) 433.
\bibitem{ueda96}
  K. Ueda, H. Kontani, M. Sigrist and P. A. Lee:  
  Phys. Rev. Lett. \textbf{76} (1996) 1932.
\bibitem{miyahara99}
  S. Miyahara and K. Ueda:
  Phys. Rev. Lett. \textbf{82} (1999) 3701.
\bibitem{taniguchi95}
  S. Taniguchi, T. Nishikawa, Y. Yasui, Y. Kobayashi, M. Sato,
  T. Nishioka, M. Kontani and K. Sano:
  J. Phys. Soc. Jpn. \textbf{64} (1995) 2758.
\bibitem{kageyama99}
  H. Kageyama, K. Yoshimura, R. Stern, N. V. Mushnikov, 
  K. Onizuka, M. Kato, K. Kosuge, C.P. Slichter, T. Goto and Y. Ueda:
  Phys. Rev. Lett. {\bf 82} (1999) 3168.
\bibitem{miyahara03}
  S. Miyahara and K. Ueda:
  J. Phys.: Condens. Matter {\bf 15} (2003) R327.
\bibitem{lauhili}
  A. L$\ddot{\rm a}$uchli, J. C. Domenge, C. Lhuillier, P. Sindzingre, M. Troyer: 
  Phys. Rev. Lett. {\bf 95} (2005) 137206.
\bibitem{nic}
  N. Shannon, T. Momoi, P. Sindzingre:
  Phys. Rev. Lett. {\bf 96} (2006) 027213.
\bibitem{wessel05}
  S. Wessel and M. Troyer: 
  Phys. Rev. Lett. \textbf{95} (2005) 127205.
\bibitem{heidarian05}
  D. Heidarian and K. Damle: 
  Phys. Rev. Lett. \textbf{95} (2005) 127206.
\bibitem{melko05}
  R. G. Melko, A. Paramekanti, A. A. Burkov, A. Vishwanath,
  D. N. Sheng and L. Balents: 
  Phys. Rev. Lett. \textbf{95} (2005) 127207.
\bibitem{mila08} 
  K. P. Schmidt, J. Dorier, A. M. L$\ddot{\rm a}$uchli, and F. Mila: 
  Phys. Rev. Lett. \textbf{100} (2008) 090401. 
\bibitem{hotta06} 
  C. Hotta and N. Furukawa: Phys. Rev. B \textbf{74} (2006) 193107.
\bibitem{hotta06-2} 
  C. Hotta, N. Furukawa, A. Nakagawa and K. Kubo: 
  J. Phys. Soc. Jpn. \textbf{75} (2006) 123704.
\bibitem{sawano05}
  F. Sawano, I. Terasaki, H. Mori, T. Mori, M. Watanabe, N. Ikeda, Y. Nogami and Y. Noda: Nature \textbf{437} (2005) 522.
\bibitem{mori98}
  H. Mori and S. Tanaka: 
  Phys. Rev. B \textbf{57} (1998) 12023.
\bibitem{yamaguchi}
  Y. Takahide, T. Konoike, K. Enomoto, M. Nishimura, T. Terashima, S. Uji, and H. M. Yamamoto: 
  Phys. Rev. Lett. \textbf{98} (2007) 116602.
\bibitem{seo04}
  For a review, see H. Seo, C. Hotta and H. Fukuyama: 
  Chem. Rev. \textbf{104} (2004) 5005. 
\bibitem{merino05}
  J. Merino, H. Seo and M. Ogata: 
  Phys. Rev. B \textbf{71} (2005) 125111.
\bibitem{kaneko06} 
  M. Kaneko and M. Ogata:
  J. Phys. Soc. Jpn. \textbf{75} (2006) 014710.
\bibitem{watanabe06} 
  H. Watanabe and M. Ogata:
  J. Phys. Soc. Jpn. \textbf{75} (2006) 063702. 
\bibitem{nishimoto08} S. Nishimoto, M. Shingai, Y. Ohta:
   arXiv:0803.0516
\bibitem{Motrunich04}
  O. I. Motrunich and P. A. Lee:
  Phys. Rev. B \textbf{69} (2004) 214516.
\bibitem{jastrow55}
  R. Jastrow: 
  Phys. Rev. \textbf{98} (1955) 1479.
%*%*%*%*%*%
\bibitem{miyazaki} The estimation of the transition point is done by the finite scaling analysis 
as shown in Fig. 3(b). The transition takes place within $V_c=12-13$. The large error bar is 
due to the fitting errors originating from the several choices of functional form we adopted. 
%*%*%*%*%*%
\bibitem{udagawa07} 
  M. Udagawa and Y. Motome: 
  Phys. Rev. Lett. \textbf{98} (2007) 206405.  
\bibitem{yokoyama06} 
  H. Yokoyama, M. Ogata and Y. Tanaka: 
  J. Phys. Soc. Jpn. \textbf{75} (2006) 114706.
\bibitem{coldea07}
 E. Wawrzy$\acute{\rm n}$ska, R. Coldea, E. M. Wheeler, I. I. Mazin, M. D. Johannes, T. S$\ddot{\rm o}$rgel, M. Jansen, R. M. Ibberson, and P. G. Radaelli: 
 Phys. Rev. Lett. \textbf{99} (2007) 157204; 
 E. Wawrzy$\acute{\rm n}$ska, R. Coldea, E. M. Wheeler, T. S$\ddot{\rm o}$rgel, M. Jansen, R. M. Ibberson, P. G. Radaelli, and M. M. Koza:  Phys. Rev. B \textbf{77} (2008) 094439. 
\end{thebibliography}
\end{document}